\documentclass[%
 reprint,
 amsmath,amssymb,
 aps,
]{revtex4-2}

\usepackage{graphicx}
\usepackage{dcolumn}
\usepackage{bm}
\usepackage{xcolor}
\usepackage{color}
\usepackage{mathtools}
\usepackage[dvipdfmx, colorlinks=true, linkcolor=red, citecolor=blue, urlcolor=blue]{hyperref}

\begin{document}



\preprint{APS/123-QED}

\title{Designing single and degenerate flat bands 
in the kagome lattice with long-range hopping}

\author{Yuta Taguchi$^{1}$}
 \email{taguchi.y@stat.phys.titech.ac.jp}
 \author{Motoaki Hirayama$^{2,3}$}%
 \author{Takashi Miyake$^{4,1}$}%
\affiliation{$^{1}$Department of Physics, Institute of Science Tokyo, 2-12-1 Oh-okayama, Meguro-ku, Tokyo 152-8551, Japan\\
$^{2}$RIKEN Center for Emergent Matter Science, Wako, Saitama, 351-0198, Japan\\
$^{3}$Department of Applied Physics, The University of Tokyo, Bunkyo-ku, Tokyo, 113-8656, Japan\\
$^{4}$CD-FMat, National Institute of Advanced Industrial Science and Technology, Tsukuba, Ibaraki 305-8568, Japan}%

\date{\today}

\begin{abstract}
We investigate the electronic structure of the kagome lattice model with first, second, and two kinds of third nearest-neighbor hoppings. 
We reveal that by tuning the third nearest-neighbor hoppings, 
not only single flat band but also degenerate flat band can be created on the $\Gamma$-M line. 
We provide the detailed conditions to realize them. 
The coexistence of these bands can be almost realized near the fundamental band gap in graphene with triangular defects in a superhoneycomb arrangement. \textcolor{black}{Furthermore, due to these flat bands, several sharp peaks appear in the optical condctivity.}
Our results strongly  indicate that long-range electron hopping has a new possibility for designing electronic structures.       
\end{abstract}

\maketitle


\section{Introduction}
Controlling the electronic structure is an important issue 
to improve materials properties of periodic systems. 
Diversity of constituent elements and structures leads to various electronic structures.
For example, flat band 
is known to significantly contribute to interesting  electronic properties such as superconductivity \cite{imada2000superconductivity, kuroki2005high, kobayashi2016superconductivity, matsumoto2018wide, mondaini2018pairing, aoki2020theoretical}, ferromegnetism \cite{mielke1991ferromagnetism, tasaki1992ferromagnetism, mielke1993ferromagnetism, kusakabe1994ferromagnetic, ramirez1994strongly, tasaki1998nagaoka, tamura2019ferromagnetism} and topological phenomena \cite{aoki1996hofstadter, vidal1998aharonov, guo2009topological, weeks2010topological, green2010isolated, tang2011high, sun2011nearly, neupert2011fractional, wang2011nearly, liu2012fractional, rhim2019classification, mizoguchi2020systematic, kuno2020extended, kuno2020interaction}. 
So far, various \textit{flat-band models} \cite{sutherland1986localization, shima1993electronic, miyahara2005flat, bergman2008band, hatsugai2011zq, misumi2017new, ramachandran2017chiral, maimaiti2019universal, mizoguchi2019flat, mizoguchi2019molecular, maimaiti2021flat, mizoguchi2021flat, mizoguchi2021flat2, mizoguchi2023unconventional, kim2023realization}, e.g., square lattice \cite{tasaki1992ferromagnetism} and kagome lattice \cite{bergman2008band, mizoguchi2019flat} have been studied. 
In general, long-range hoppings break the flatness of a band, but some research shows how to maintain a single flat band with the addition of next-nearest neighbor hopping\cite{mizoguchi2019flat, mizoguchi2023unconventional}.
Exploration of materials 
\cite{yamada2016first, hase2018possibility, maruyama2016coexistence, maruyama2017interplay, lee2019hidden, kang2020topological, kang2020dirac, fleurence2020emergence, mizoguchi2023unconventional, kim2023realization} which realize flat-band model 
is also an important issue. Recently, data science has been applied to explore flat-band materials more efficiently \cite{bhattacharya2023deep}. 

Another intriguing property in electronic structure is degenerate band. 
Degeneracy in band dispersion such as Dirac cone and nodal line has attracted much attention since they trigger non-trivial physical phenomena such as the appearance of  drumhead surface state and huge anomalous Nernst effect \cite{sakai2020iron, chen2022large}. Therefore, not only a single flat band but also a degenerate flat band is worth studying. In general, degenerate band has been studied in terms of structure symmetry \cite{young2015dirac, wieder2016double, takahashi2017spinless, yang2017topological, zhang2019catalogue}, so whether degenerate flat band can be designed without changing symmetry, is an interesting question for flat-band model and flat-band materials design.

In this \textcolor{violet}{paper}, 
we study the kagome lattice model including long-range hoppings. 
We reveal that not only a new single flat band but also a degenerate flat band appears along certain direction ($\Gamma-\rm{M}$ line in the Brillouin zone) by introducing long range hopping, i.e., the third nearest neighbor hopping, under certain conditions. We show that dispersive band in the kagome lattice with only the nearest neighbor hopping can be become flat by long-range hopping. Therefore, the flat band in our work has a different character from that in the previous works \cite{mizoguchi2019flat, mizoguchi2023unconventional}.
To the best of our knowledge, method for designing degenerate flat band has never been proposed theoretically in the kagome lattice. We also show that the above conditions are realized in graphene with triangular defects by extending the distance between defects. \textcolor{black}{In addition, we explain that these flat bands give rise to the sharp peaks in the optical conductivity of this defective graphene.}  


\begin{figure*}
\includegraphics[scale=0.85]{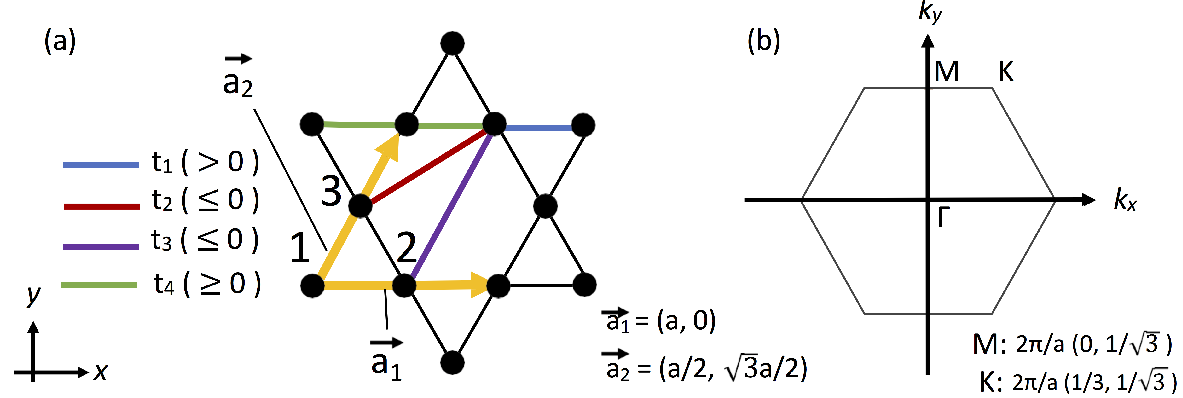}
\caption{\label{fig:fig3} (a) Schematic views of the kagome lattice. $\vec{a_1}$ and $\vec{a_2}$ are the primitive translation vectors of kagome lattice. Three sublattices (1,2,3) and four kinds of electron hoppings ($t_1,t_2$, $t_3$ and $t_4$) are shown. 
It is to be noted that $t_4$ represents electron hopping between sublattice $i$ and sublattice $i$. (b) First Brillouin zone of the kagome lattice. }
\end{figure*}

\section{Kagome lattice with the first, the second, and the third nearest neighbor hoppings}

\subsection{Model and hamiltonian}
We focus on the kagome lattice with not only the nearest neighbor electron hopping ($t_1$) but also the second ($t_2$) and the third ($t_3$ and $t_4$) nearest neighbor electron hoppings as shown in Fig.~\ref{fig:fig3}(a).

Here, let us define P as the point which internally divides the $\Gamma$-M line (Fig.~\ref{fig:fig3}(b)) into $k:(1-k)$ $(0\leqq k \leqq1)$. The Bloch Hamiltonian at the P point of this model $\mathcal{H}(k)$ can be expressed as

\begin{widetext}
\begin{equation}
\mathcal{H}(k)=\begin{pmatrix}
2t_3\cos(k\pi)+2t_4[1+\cos(k\pi)] & 2t_1+2t_2\cos(k\pi) & (t_1+t_2)(1+e^{-ik\pi}) \\
2t_1+2t_2\cos(k\pi) & 2t_3\cos(k\pi)+2t_4[1+\cos(k\pi)] & (t_1+t_2)(1+e^{-ik\pi}) \\
(t_1+t_2)(1+e^{ik\pi}) & (t_1+t_2)(1+e^{ik\pi}) & 2t_3+4t_4\cos(k\pi) \\
\end{pmatrix}.
\label{eq:eqr1}
\end{equation}
\end{widetext}
It is to be noted that the $(i,j)$ component of $\mathcal{H}(k)$ denotes the electron hopping between the sublattices $i$ and $j$ in Fig.~\ref{fig:fig3}(a). The detailed derivation of $\mathcal{H}(k)$ is shown in \textcolor{black}{Appendix~\ref{appendix:Appendix_derivation}}. 
From this $\mathcal{H}(k)$, we can obtain the Hamiltonians at the M point ($\mathcal{H}_{\rm{M}}$), the middle point of the $\Gamma$-M line ($\mathcal{H}_{\rm{middle}}$) and the $\Gamma$ point ($\mathcal{H}_\Gamma$).

We use these three Hamiltonians to study the conditions of $(t_1, t_2, t_3, t_4)$ which realize 
single flat band or degenerate flat bands on the $\Gamma$-M line hereafter.


\begin{figure*}
\includegraphics[scale=0.85]{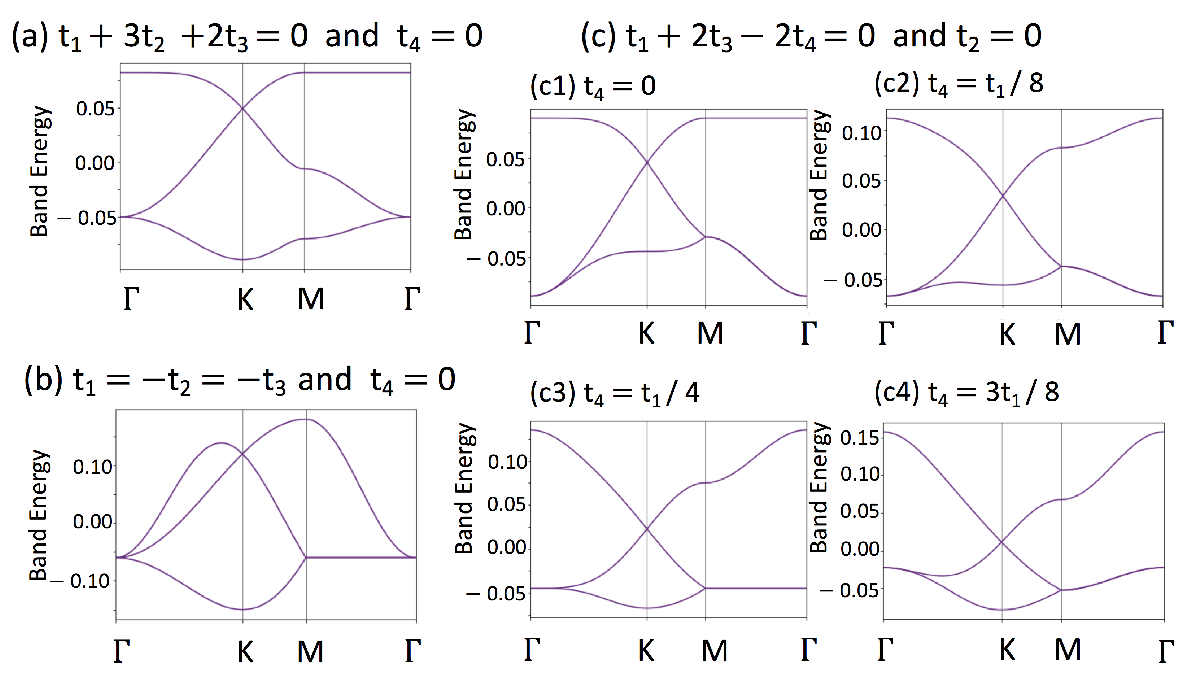}
\caption{\label{fig:fig1} Electronic structures of kagome lattice when $(t_1, t_2, t_3, t_4)$ satisfies (a) $t_1+3t_2+2t_3=0$ and $t_4=0$, (b) $t_1=-t_2=-t_3$ and $t_4=0$, and (c) $t_1+2t_3-2t_4=0$ and $t_2=0$, respectively. We calculate these electronic structures using the PythTB package \cite{Pyth}.}
\end{figure*}

\subsection{Electronic structure}

Figure~\ref{fig:fig1} (a) shows the electronic structure for  $t_1 + 3 t_2 + 2 t_3 = 0$ and $t_4=0$, obtained by diagonalizing eq.~(\ref{eq:eqr_hamiltonian}). 
We see that a single flat band appears between M and $\Gamma$. 
We also find another condition to realize flat band. 
When $t_1 = -t_2 = -t_3$ and $t_4 = 0$ are satisfied, degenerate flat band appears along the $\Gamma-\rm{M}$ line, 
as shown in Fig.~\ref{fig:fig1}(b). 
It is to be noted that degenerate flat band cannot be designed when only $t_1$ and $t_2$ are considered. This result strongly indicates that long-range electron hopping plays an important role in creating various types of electronic structures. In addition, the degenerate flat band by using long range hopping is realized without changing system symmetry. 

Figure~\ref{fig:fig1}(c1) is the result for the special case of (a), i.e., 
$t_1 + 3 t_2 + 2 t_3 = 0$ and $t_2 = t_4 = 0$. 
We find a single flat band and a degenerate (but not flat) band along the $\Gamma-\rm{M}$ line.
As we increase $t_4$ with keeping $t_1 + 2 t_3 - 2 t_4 = 0$, the width of the degenerate band decreases, and 
becomes flat when $t_4 = t_1 / 4$, then turns to increase (Fig.~\ref{fig:fig1}(c2)-(c4)).

To understand the above results, we first focus on the case of $t_4=0$. 
The conditions that the flat band appears can be derived by focusing on the eigenvalues of $\mathcal{H}_{\rm{M}}$, $\mathcal{H}_{\rm{middle}}$ and $\mathcal{H}_{\Gamma}$. The detailed derivations are given in \textcolor{black}{Appendices~\ref{appendix:AppendixB} and ~\ref{appendix:AppendixB2}}. The single flat band along the $\Gamma-\rm{M}$ line in Fig.~\ref{fig:fig1}(a) is dispersive in the case of kagome lattice with only $t_1$ as shown in \textcolor{black}{Appendix~\ref{appendix:AppendixA}}. By introducing $t_2$ and $t_3$ appropriately, this dispersive band loses dispersion along the $\Gamma-\rm{M}$ line. 
Appearing the degenerate flat band along the $\Gamma-\rm{M}$ line in Fig,~\ref{fig:fig1}(b) is interesting, but it is hard to realize in real materials due to the condition of $t_1=-t_2=-t_3$. It would be very difficult to have the same absolute value for $t_1$ and $t_3$. However, by also introducing $t_4$, a degenerate flat band along the $\Gamma-\rm{M}$ line can be designed under more realistic condition, as will be discussed later.

Secondly, we focus on the kagome lattice considering $t_1,t_2,t_3$ and $t_4$. As shown in Fig.~\ref{fig:fig1}(c) and Fig.~\ref{fig:figAppendixD}, a degenerate band always appears along the $\Gamma-\rm{M}$ line under the condition of (i) $t_1+2t_3-2t_4=0$ and $t_2=0$, or (ii) $t_1=-t_2$ and $t_4-t_3=t_1$. The detailed derivations of these condition are shown in \textcolor{black}{Appendix~\ref{appendix:AppendixC}}. In this section, we discuss (i) $t_1+2t_3-2t_4=0$ and $t_2=0$ case. 
When $t_4=\alpha t_1$, the eigenvalue of the degenerate band is $[(2\alpha-2)+(4\alpha-1)\cos(k\pi)]t_1$ ($0\leq k\leq1$, \textcolor{black}{see Appendix~\ref{appendix:AppendixC2}}), so the width of the degenerate band is $|(2-8\alpha)t_1|$. Therefore, this bandwidth monotonically decreases with $\alpha$ ($0\leq\alpha\leq \frac{1}{4}$) and increases with $\alpha$ ($\frac{1}{4}\leq\alpha$). When $\alpha=\frac{1}{4}$, the degenerate band becomes flat as shown in Fig.~\ref{fig:fig1}(c3). Hence, $t_4$ smaller than $\frac{t_1}{4}$ plays a crucial role in flattening the degenerate band along the $\Gamma-\rm{M}$ line. The condition for degenerate flat band is $(t_1,t_2,t_3,t_4)=(1, 0, -\frac{1}{4}, \frac{1}{4})t_1$ (Fig.~\ref{fig:fig1}(c3)), which is more realistic than $t_1=-t_2=-t_3$ (Fig.~\ref{fig:fig1}(b)).


\begin{figure}
\includegraphics[scale=0.40]{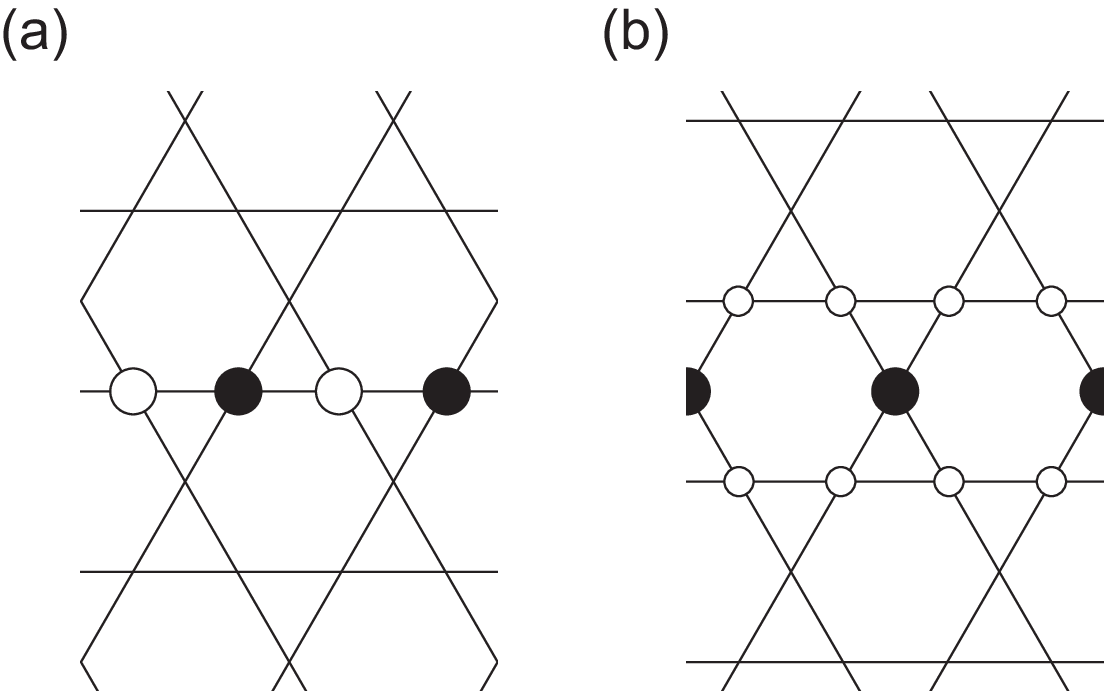}
\caption{\label{fig:fig4} Localized polymer orbitals at degenerate flat bands (Fig.~\ref{fig:fig1} (c3)). Black and white circles correspond to positive and negative phases, respectively. The coefficients of these orbitals are (a) $\frac{1}{\sqrt{2}}$ (black) and $-\frac{1}{\sqrt{2}}$ (white), and (b) $\frac{2}{\sqrt{5}}$ (black) and $-\frac{1}{2\sqrt{5}}$ (white), respectively. }
\end{figure}

\begin{figure*}
\includegraphics[scale=0.95]{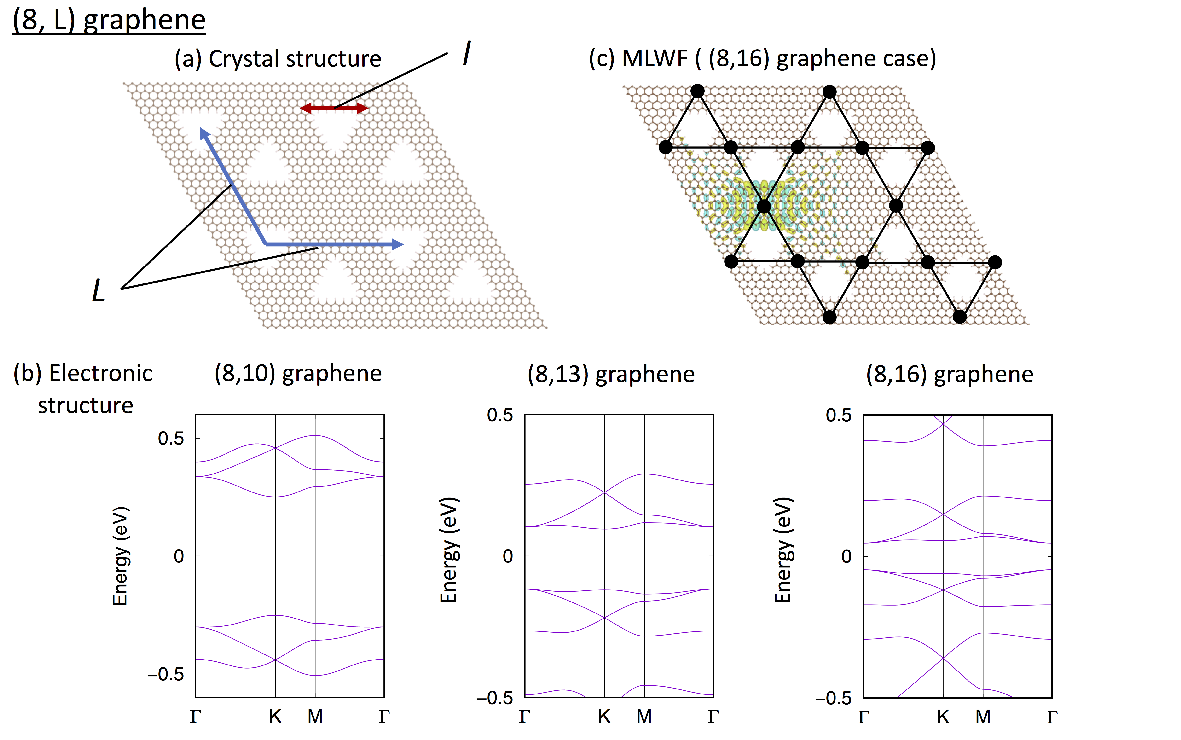}
\caption{\label{fig:fig2} (a) Crystal structure, (b) electronic structure of the (8,10), (8,13), and (8,16) graphene systems, and (c) one of the obtained maximally localized Wannier functions (MLWFs) of the $(8,16)$ graphene system. An integer $l$ is defined as the defect side length in units of the pristine graphene lattice constant. An integer $L$ is defined as the length of primitive translation vectors (two blue arrows) in units of the pristine graphene lattice constant. Brown and pink balls represent carbon and hydrogen atoms, respectively. VESTA \cite{momma2008vesta} was used to visualize the crystal structure. In the electronic structure, the middle of the CBM (conduction band minimum) and the VBM (valence band maximum) is set to zero. In the MLWF, sign differences are shown by two colors. To obtain the electronic structure, we performed first-principles total-energy calculations within the framework of the density functional theory (DFT).
The electronic structure and the MLWFs are obtained by using the QuantumESPRESSO package \cite{giannozzi2009quantum}. }
\end{figure*}


\subsection{Real space picture of degenerate flat band}

Here, let us discuss the interference effect and the real space picture in the case of Fig.~\ref{fig:fig1}(2c) where the degenerate flat band along the $\Gamma-\rm{M}$ line appears. Firstly, in the case of the kagome-lattice model with only $t_1$, any kinds of Kohn-Sham wave function in the flat band can be constructed by linear combination of localized molecular orbitals. This molecular orbital is localized in real space due to quantum interference effect of hopping and phases of atomic orbitals. Next, in our present study, degenerate flat band appears along the $\Gamma-\rm{M}$ line as shown in Fig.~\ref{fig:fig1}(2c). Similar to the kagome-lattice model with only $t_1$, we can design any kinds of Kohn-Sham wave function along the $\Gamma-\rm{M}$ line by linear combination of localized doubly degenerate polymer orbitals. These polymer orbitals are shown in Fig.~\ref{fig:fig4}. The polymer orbital shown in Fig.~\ref{fig:fig4}(a), is the eigenstate in the case of the kagome-lattice model with only $t_1$, and the linear combination of it constructs a flat band along the $\Gamma-\rm{M}$ line. This orbital is also the eigenstate when we consider the long range hopping $t_3$ and $t_4$ due to interference effect of them. On the other hand, the orbital shown in Fig.~\ref{fig:fig4}(b), is not the eigenstate in the case of the kagome-lattice model with only $t_1$. When the state at the $\Gamma$ point is designed by the linear combination of this orbital, this state can be an eigenstate degenerated with the polymer orbital shown in Fig.~\ref{fig:fig4}(a), but degeneracy does not occur at the general point on the $\Gamma-\rm{M}$ line and the energy dispersion is not flat. In the case of the kagome-lattice model with $t_1$, $t_3$ and $t_4$, the polymer orbital shown in Fig.~\ref{fig:fig4}(b) is the eigenstate and it has the same energy ($-\frac{3t_1}{2}$) as the polymer orbital shown in Fig.~\ref{fig:fig4}(a).
Considering the fact that the irreducible representations of these wavefunctions of the polymers are different (Fig.~\ref{fig:fig4}), the degeneracy along the $\Gamma-\rm{M}$ line in Fig.~\ref{fig:fig1}(2c) is not protected by two dimensional representation.
We find that two properties, degeneracy and localization, are realized by interference effect of electron hopping.

\subsection{Effect of small perpendicular hopping on electronic structure of Kagome lattice}
So far, we focus on two dimensional kagome lattice. However, in the real material cases, small perpendicular hopping exists. In order to unveil how small perpendicular hopping affects the flat dispersion, we performed additional calculations for 3D systems including perpendicular hopping. When perpendicular hopping is a tenth of the nearest neighbor hopping, the flat dispersion is sometimes broken, but this bandwidth is very small as shown in Fig.~\ref{fig:figSprb}(b1). \textcolor{black}{In Appendix~\ref{appendix:AppendixE}}, we show the detailed discussion and results.



\begin{table}[b]
\caption{\label{tab:tabprb1}%
 In the case of $(8,L)$ graphenes, (i) Bandwidth of the first LU band ($W_{\rm{1st\,LU}}$), the second LU band ($W_{\rm{2nd\, LU}}$), and the third LU band ($W_{\rm{3rd\,LU}}$) along the $\Gamma-\rm{M}$ line, and (ii) Energy difference between the first and the second LU bands ($\epsilon_{\rm{dif,\,M}}$) at the $\rm{M}$ point.}
\begin{ruledtabular}  
\begin{tabular}{ccccc}
\textrm{$(l,L)$}&
\multicolumn{1}{c}{\textrm{$W_{\rm{1st\,LU}}$ (eV)}}&
\multicolumn{1}{c}{\textrm{$W_{\rm{2nd\,LU}}$ (eV)}}&
\multicolumn{1}{c}{\textrm{$W_{\rm{3rd\,LU}}$} (eV)}&
\multicolumn{1}{c}{\textrm{$\epsilon_{\rm{dif,\,M}}$} (eV)}\\
\colrule
(8,10) &\multicolumn{1}{c}{0.043} &\multicolumn{1}{c}{0.031} &\multicolumn{1}{c}{0.11} &\multicolumn{1}{c}{0.073}\\
(8,13) & \multicolumn{1}{c}{0.015} &\multicolumn{1}{c}{0.042} &\multicolumn{1}{c}{0.036} &\multicolumn{1}{c}{0.028}\\
(8,16) & \multicolumn{1}{c}{0.024} &\multicolumn{1}{c}{0.034} &\multicolumn{1}{c}{0.016} &\multicolumn{1}{c}{0.010}\\
\end{tabular}
\end{ruledtabular}
\end{table}


\section{Material realization}

Next, we discuss materials realization of the flat band. 
The electronic structure of the kagome-lattice model with $(t_1,t_2,t_3,t_4)$, with $t_1+2t_3-2t_4=0$ and $t_2=0$ (Fig.~\ref{fig:fig1}(c)), can be nearly realized in graphene with triangular defects in a superhoneycomb arrangement, which is called $(l,L)$ graphene in our previous paper \cite{taguchi2023electronic} (Fig~\ref{fig:fig2}(a)). 
In the caption of Fig~\ref{fig:fig2}, the definitions of $l$ and $L$ are explained. It is to be noted that experimental realization of $(l,L)$ graphene has been a difficult issue. One of the reasons is that interdefect distance is so small ($\thicksim$ nm). To the best of our knowledge, there is no experimental technique for making defects periodically with this pitch although many kinds of techniques such as electron-beam lithography \cite{eroms2009weak, giesbers2012charge, sandner2015ballistic} and block copolymer lithography \cite{bai2010graphene, kim2010fabrication, kim2012electronic} have been developed. Recently, however, by using helium ion beam milling, Schmidt \textit{et al.} and Liu \textit{et al.} succeeded in creating defects with a pitch of 18 nm and 15 nm, respectively \cite{schmidt2018structurally, liu2020conductance}. If technique of creating defects is further improved, graphene having defects with a pitch of several nanometers, including the $(l,L)$ graphene, may be realized. Another promising method for synthesis of $(l,L)$ graphene is the use of a hexagonal boron nitride (\textit{h}-BN) layer, which has a honeycomb lattice structure consisting of alternating boron and nitrogen toms, and platinum as a catalyst. In 2009, by using an electron beam, Jin \textit{et al.} succeeded in resolving triangular defects in \textit{h}-BN monolayer, keeping distance between defects equals to several nanometers \cite{jin2009fabrication}. Furthermore, in 2020, Kim \textit{et al.} reported the conversion of \textit{h}-BN to graphene on platinum substrate when \textit{h}-BN was exposed to methane gas \cite{kim2020effect}. Combining these experimental results, graphene with triangular defects with a pitch of several nanometers may be realized in the future. Therefore, $(l,L)$ graphene is worth studying theoretically in detail. 

\subsection{Electronic structure of $(8,L)$ graphene}
In Fig~\ref{fig:fig2}(b), the electronic structures of $(8,10)$, $(8,13)$ and $(8,16)$ graphenes obtained by the first-principle calculations are shown, respectively. We show the details of the \textit{ab initio} calculation \textcolor{black}{in the Appendix~\ref{appendix:Appendix_method_DFT}.}
In the discussion that follows, we focus on the first lowest-unoccupied (LU) and the second LU and the third LU bands. As $L$ increases, the shape of these bands becomes similar to that shown in Fig.~\ref{fig:fig1}(c1). Actually, the third LU bandwidth along the $\Gamma-\rm{M}$ line ($W_{\rm{3rd\,LU}}$ in Table.~\ref{tab:tabprb1}) is monotonically decreases with $L$. In addition, the energy difference between the first and the second LU band at the $\rm{M}$ point ($\epsilon_{\rm{dif,\,M}}$ in the Table.~\ref{tab:tabprb1}) also monotonically decreases with $L$, and $\epsilon_{\rm{dif,\,M}}$ of the $(8,16)$ graphene is 0.010 eV, indicating that the first and the second LU bands almost degenerate along the $\Gamma-\rm{M}$ line (Fig.~\ref{fig:fig2}(b)).  Hence, we expect that the kagome-lattice model with $(t_1,t_2,t_3,t_4)$, where $t_1+2t_3-2t_4=0$ and $t_2=0$, is nearly realized when $L$ is large. Actually, as shown below, we can confirm it by constructing the maximally localized Wannier functions (MLWFs) from these bands.

\subsection{Maximally localized Wannier function of $(8,L)$ graphene}

The WLWF has a Wannier center in the center of the graphene nanoribbon, forming a kagome lattice(Fig~\ref{fig:fig2}(c) and \textcolor{black}{Fig.~\ref{fig:figS2}(b)}).
The MLWF originates from the hybridized two edge states of the nanoribbon.
In Table~\ref{tab:table1}, the calculated $(t_1,t_2,t_3,t_4)$ of $(8,L)$ graphenes ($L=10,13$ and 16) are shown.
Interestingly, the value of $|t_3|$ does not decrease with increasing $L$ while the value of $|t_2|$ decreases sharply with $L$.
The MLWF extends outside of the subribbon part, and due to this, $t_3$ value is nonzero for large $L$. 
We find that $t_2\simeq0$, $t_3\simeq-\frac{1}{2}t_1$, and $t_4\simeq0$ results in $t_1+2t_3-2t_4\simeq0$ and $t_2\simeq0$ for large $L$ case. 

\begin{table}[h]
\caption{\label{tab:table1}%
 $(t_1,t_2,t_3,t_4)$ and the value of $t_1+2t_3-2t_4$ in the case of $(8,L)$ graphenes.}
\begin{ruledtabular}  
\begin{tabular}{cccccc}
\textrm{$L$}&
\multicolumn{1}{c}{\textrm{$t_1$ (eV)}}&
\multicolumn{1}{c}{\textrm{$t_2$ (eV)}}&
\multicolumn{1}{c}{\textrm{$t_3$} (eV)}&
\multicolumn{1}{c}{\textrm{$t_4$} (eV)}&
\multicolumn{1}{c}{\textrm{$t_1+2t_3-2t_4$} (eV)}\\
\colrule
10 &\multicolumn{1}{c}{0.031} &\multicolumn{1}{c}{-0.021} &\multicolumn{1}{c}{-0.0099} &\multicolumn{1}{c}{0.00038}&\multicolumn{1}{c}{0.010}\\
13 & \multicolumn{1}{c}{0.033} &\multicolumn{1}{c}{-0.0069} &\multicolumn{1}{c}{-0.013} &\multicolumn{1}{c}{0.0018}&\multicolumn{1}{c}{0.0034}\\
16 & \multicolumn{1}{c}{0.029} &\multicolumn{1}{c}{-0.0028} &\multicolumn{1}{c}{-0.013} &\multicolumn{1}{c}{0.0028}&\multicolumn{1}{c}{-0.0026}\\
\end{tabular}
\end{ruledtabular}
\end{table}

Let us focus on the electronic structure of $(8,16)$ graphene shown in Fig.~\ref{fig:fig2}.
On the $\Gamma-\rm{M}$ line, the nearly single flat band (bandwidth: 0.016 eV) and the degenerate flat band (bandwidth: 0.024 eV and 0.034 eV) coexist. In the case of $(8,16)$ graphene, $t_4\simeq\frac{1}{10}t_1(<\frac{1}{4}t_1)$ (Table~\ref{tab:table1}) holds, so $t_4$ plays a role in decreasing the degenerate bandwidth. Actually, if we do not consider $t_4$, the degenerate bandwidth is $2t_1=0.058$ eV, which is larger than that of the first-principles calculation. On the other hand, $t_4$ plays a role in increasing the single bandwidth ($=|8t_4|$). However, due to very small $t_4$ value ($t_4=-0.0028$ eV (Table.~\ref{tab:table1})), the single band is nearly flat.

\subsection{Optical conductivity of $(8,L)$ graphene}
\begin{figure*}
\includegraphics[scale=0.85]{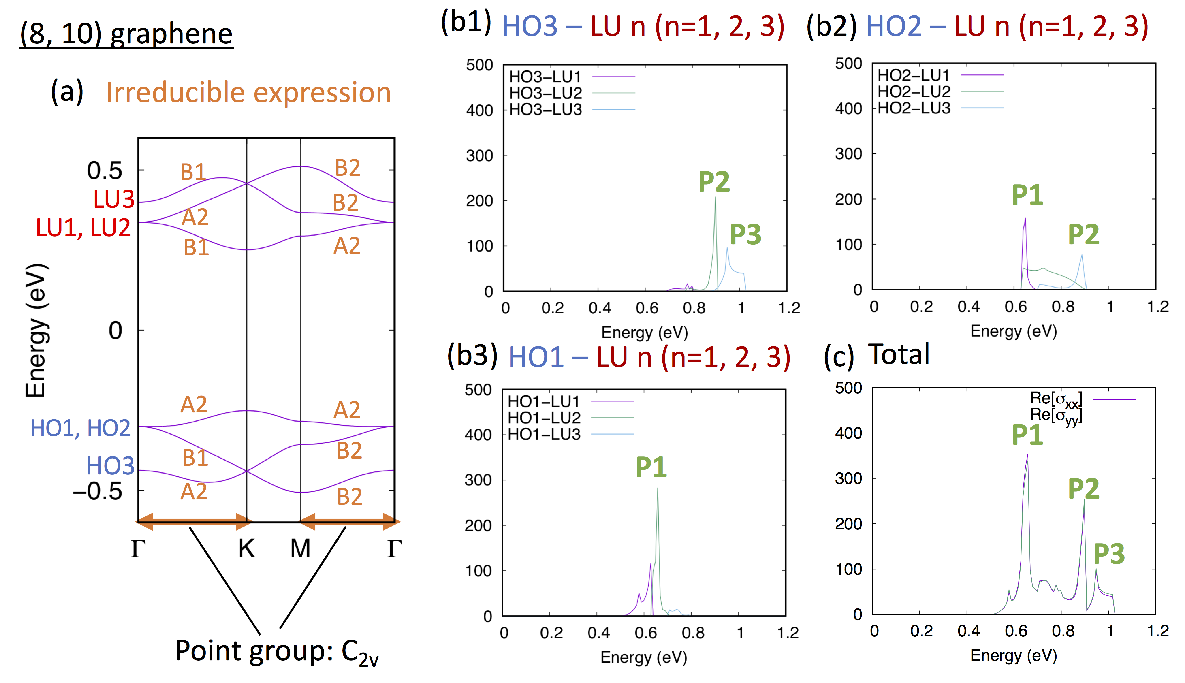}
\caption{\label{fig:figS8} (a) Irreducible expression along the $\Gamma-\rm{K}$ and $\rm{M}-\Gamma$ lines, (b1), (b2) and (b3) partial optical conductivities, and (c) Total optical conductivity of $(8,10)$ graphene.}
\end{figure*}

\begin{figure*}
\includegraphics[scale=0.85]{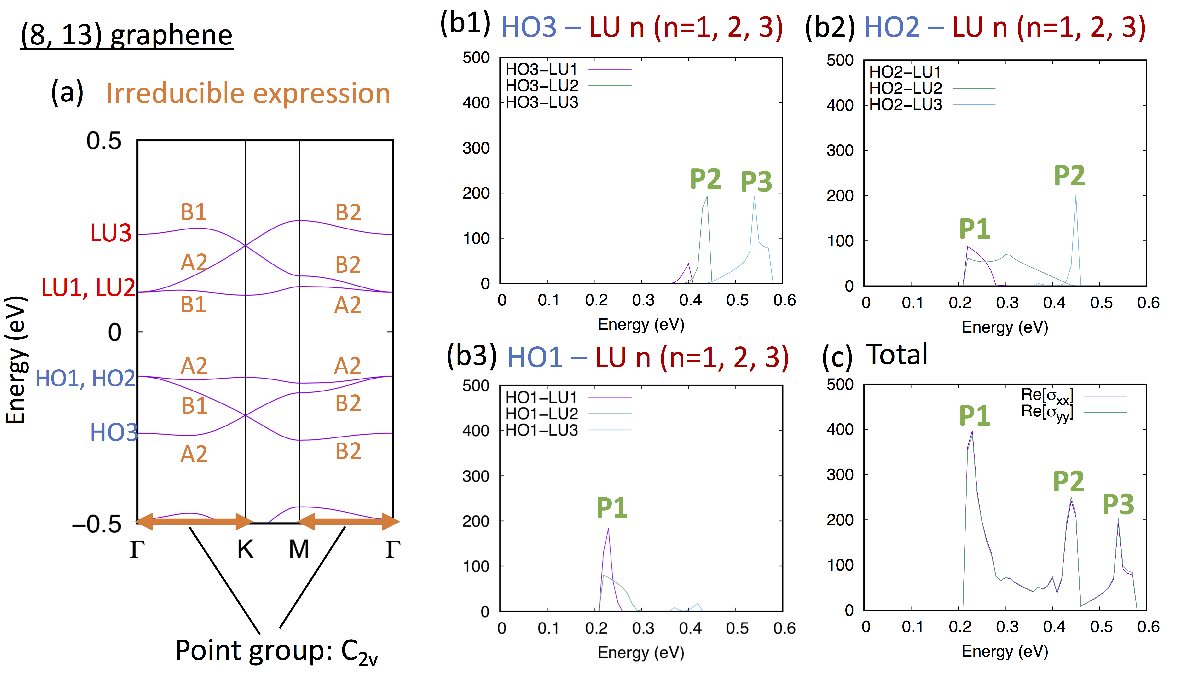}
\caption{\label{fig:figS4} (a) Irreducible expression along the $\Gamma-\rm{K}$ and $\rm{M}-\Gamma$ lines, (b1), (b2) and (b3) partial optical conductivities, and (c) Total optical conductivity of $(8,13)$ graphene.}
\end{figure*}

\begin{figure*}
\includegraphics[scale=0.85]{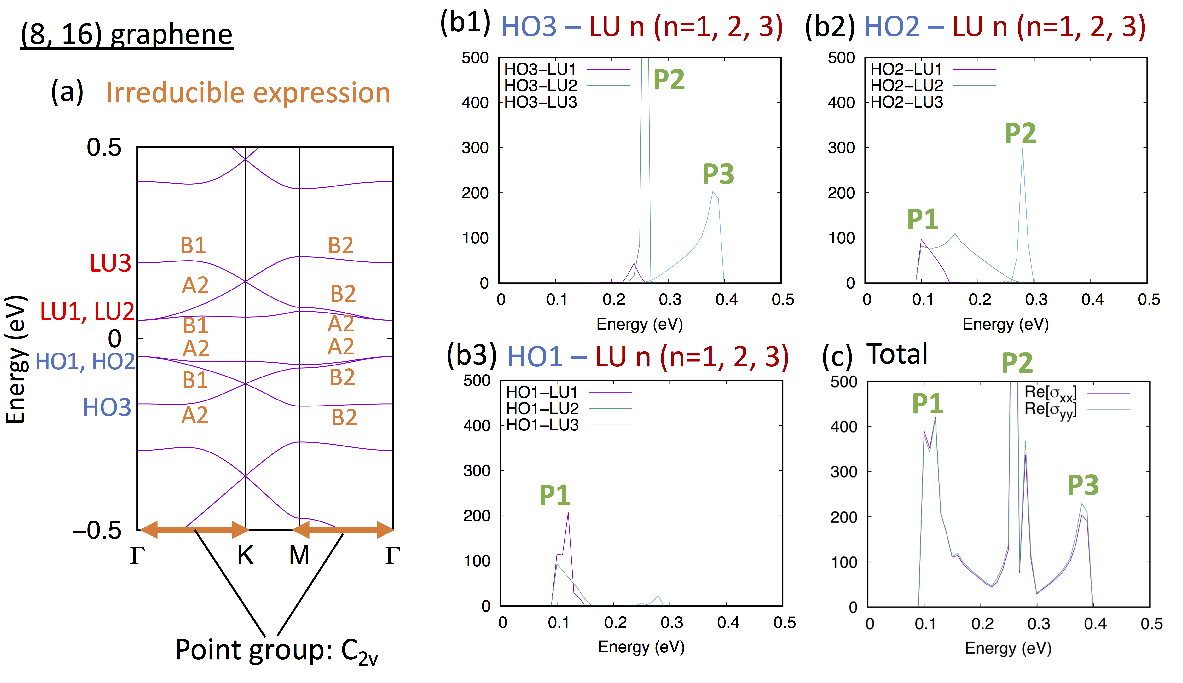}
\caption{\label{fig:figS5} (a) Irreducible expression along the $\Gamma-\rm{K}$ and $\rm{M}-\Gamma$ lines, (b1), (b2) and (b3) partial optical conductivities, and (c) Total optical conductivity of $(8,16)$ graphene.}
\end{figure*}

Finally, we also investigate how these nearly flatness along the $\Gamma-\rm{M}$ line affect experimental observables. In this study, we focus on the optical conductivity. \textcolor{black}{The computational details are shown in Appendix~\ref{appendix:Appendix_optical}}.

\textcolor{black}{In Figs.~\ref{fig:figS8}, \ref{fig:figS4} and \ref{fig:figS5}, the irreducible expression along the $\Gamma-\rm{K}$ and $\Gamma-\rm{M}$ lines (Figs.~\ref{fig:figS8}(a), ~\ref{fig:figS4}(a) and \ref{fig:figS5}(a)), partial optical conductivity (Figs.~\ref{fig:figS8}(b1-b3), ~\ref{fig:figS4}(b1-b3) and \ref{fig:figS5}(b1-b3)) and total optical conductivity (Figs.~\ref{fig:figS8}(c), ~\ref{fig:figS4}(c) and \ref{fig:figS5}(c)) of $(8,10)$, $(8,13)$ and $(8,16)$ graphene are shown. In the case of the (8,16) graphene, which has the nearly single flat band and the nearly degenerate flat band along the $\Gamma-\rm{M}$ line in the electronic structure, three sharp peaks ($\rm{P}_{1}$, $\rm{P}_{2}$ and $\rm{P}_{3}$) appear as shown in Fig.~\ref{fig:figS5}(c). To reveal the origin of these three peaks, we calculate partial optical conductivity and irreducible expressions of the first, second and third LU(HO) bands as shown in Figs.~\ref{fig:figS5}(a) and ~\ref{fig:figS5}(b). 
It is to be noted that the partial optical conductivity represents optical conductivity by transition between the $m$th HO and the $n$th LU bands. }

\textcolor{black}{Let us discuss the origin of $\rm{P}_{1}$, $\rm{P}_{2}$, and $\rm{P}_{3}$. Firstly, as shown in the results of partial optical conductivities, the origin of $\rm{P}_{1}$ is the transition between (i) the first HO and the first LU bands, (ii) the first HO and the second LU bands, (iii) the second HO and the first LU bands, and (iv) the second HO and the second HO bands. Among these four types of transition, transition between (i) generates the largest optical conductivity, as shown in Fig.~\ref{fig:figS5}(b). Next, considering the irreducible expression, the transitions between (ii) and between (iii) should be derived from the nearly degenerate flat bands along the $\Gamma-\rm{M}$ line. From Figs.~\ref{fig:figS5}(b2) and ~\ref{fig:figS5}(b3), it is found that these transitions generate the almost same optical conductivity as that of the transition between (i). Therefore, the nearly degenerate flat band along the $\Gamma-\rm{M}$ line has an important role in creating $\rm{P}_{1}$.}

\textcolor{black}{Secondly, the origin of $\rm{P}_{2}$ is the transition between the second HO (LU) and the third LU (HO) bands. In terms of energy range of $\rm{P}_{2}$ and irreducible expression, these transition only occurrs along the $\rm{K}-\rm{M}$ line. The second HO (LU) and the third LU(HO) bands are nearly parallel along the $\rm{K}-\rm{M}$ line, resulting in significantly sharp shape of $\rm{P}_{2}$.}

\textcolor{black}{Thirdly, the origin of $\rm{P}_{3}$ is the transition between the third HO and the third LU bands. In terms of energy range of $\rm{P}_{3}$ and irreducible expression, this transition only occurs along the $\Gamma-\rm{K}$ line, indicating that nearly single flat parts of the third HO and LU bands along the $\Gamma-\rm{K}$ line, generate $\rm{P}_{3}$. Therefore, nearly single flat band is found to create new peak, $\rm{P}_{3}$. The $\rm{P}_{3}$, originated from the transition between the third HO and the third LU bands, can be also seen in the case of $(8,13)$ graphene (Fig.~\ref{fig:figS4}(c)) since nearly single flat parts of the third HO and LU bands along the $\Gamma-\rm{K}$ line appear, as shown in Fig.~\ref{fig:figS4}(a). }


\section{Summary}
In this \textcolor{black}{paper}, we have studied the electronic structure of the kagome lattice with the first ($t_1$), the second ($t_2$) and the third ($t_3,t_4$) nearest neighbor hoppings. It is found that not only single flat band but also degenerate flat band on the $\Gamma-\rm{M}$ line can be designed by introducing $t_3$ and $t_4$ appropriately. It is realized in graphene with triangular defects in a superhoneycomb arrangement. By extending distance between defects, $(t_1,t_2,t_3,t_4)$ approaches the condition for appearing degenerate flat band, and then, nearly single flat and nearly degenerate flat band coexist on the $\Gamma-\rm{M}$ line. \textcolor{black}{These nearly flat bands play an important role in creating the sharp peaks of optical conductivity.}

\begin{acknowledgments}
YT acknowledge the financial support from JST SPRING, Grant Number JPMJSP2106, and Tokyo Tech Academy for Convergence of Materials and Informatics (TAC-MI). Numerical calculations were done at the Supercomputer Center of the Institute for Solid State Physics, The University at Tokyo; at the Center for Computational Materials Science, Institute for Materials Research, Tohoku University for the use of MASAMUNE-IER; and at the Global Scientific Information and Computing Center of the Tokyo Institute of Technology. 
\end{acknowledgments}

\nocite{*}


\appendix
\begin{widetext}
\section{Kagome lattice with only the nearest neighbor hopping ($t_1$)}
\label{appendix:AppendixA}
\begin{figure}[h]
\includegraphics[scale=0.95]{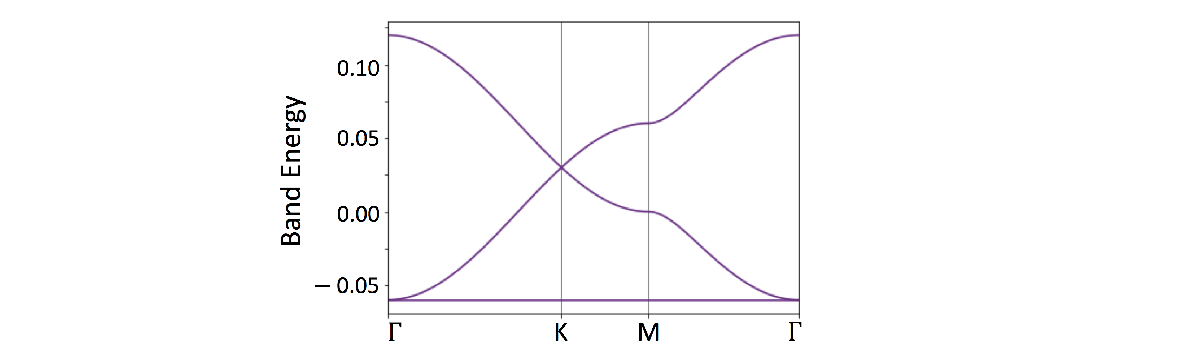}
\caption{\label{fig:AA1} \textcolor{black}{Electronic structure of kagome lattice with $t_1=0.03$ eV, $t_2=t_3=t_4=0.00$ eV.} }
\end{figure}

\textcolor{black}{
When the values of $t_2$, $t_3$ and $t_4$ are 0, $\mathcal{H}(\bm{k})$ (Eq.~\ref{eq:eqr_hamiltonian}) can be given as follows.}

\textcolor{black}{
\begin{equation}
\mathcal{H}(\bm{k})=\begin{pmatrix}
0 & t_1+ t_1e^{-i\bm{a_1}\cdot\bm{k}} & t_1+ t_1e^{-i\bm{a_2}\cdot\bm{k}} \\
t_1+ t_1e^{i\bm{a_1}\cdot\bm{k}} & 0 & t_1+ t_1e^{i(\bm{a_1}-\bm{a_2})\cdot\bm{k}} \\
t_1+ t_1e^{i\bm{a_2}\cdot\bm{k}} & t_1+ t_1e^{-i(\bm{a_1}-\bm{a_2})\cdot\bm{k}} & 0 \\
\end{pmatrix}.
\end{equation}
From this $\mathcal{H}(\bm{k})$, the electronic structure can be obtained as shown in Fig.~\ref{fig:AA1}. The lowest band is completely flat while the second and the third lowest bands are dispersive.}

\section{Detailed derivation of $\mathcal{H}(k)$}
\label{appendix:Appendix_derivation}
The Bloch hamiltonian $\mathcal{H}(\bm{k})$ of the kagome lattice model with $t_1$, $t_2$, $t_3$ and $t_4$ ($\bm{k}$ is general wave vector) can be given as follows:

\textcolor{black}{
\begin{equation}
\mathcal{H}(\bm{k})=\begin{pmatrix}
H_{1.1}(\bm{k}) & H_{1.2}(\bm{k})& H_{1.3}(\bm{k})\\
H_{2.1}(\bm{k}) & H_{2.2}(\bm{k}) & H_{2.3}(\bm{k}) \\
H_{3.1}(\bm{k}) & H_{3.2}(\bm{k}) & H_{3.3}(\bm{k}) \\
\end{pmatrix}
\label{eq:eqr_hamiltonian}
\end{equation}}
\textcolor{violet}{with}

\textcolor{black}{
\begin{subequations}
\label{eqr2a}
\begin{eqnarray}
H_{1.1}(\bm{k})=t_3e^{i(\bm{a_1}-\bm{a_2})\cdot\bm{k}}+t_3e^{-i(\bm{a_1}-\bm{a_2})\cdot\bm{k}}+t_4e^{i\bm{a_1}\cdot\bm{k}}+t_4e^{-i\bm{a_1}\cdot\bm{k}}+t_4e^{i\bm{a_2}\cdot\bm{k}}+t_4e^{-i\bm{a_2}\cdot\bm{k}},
\end{eqnarray}
\begin{eqnarray}
H_{1.2}(\bm{k})= t_1+ t_1e^{-i\bm{a_1}\cdot\bm{k}}+t_2e^{-i\bm{a_2}\cdot\bm{k}}+t_2e^{-i(\bm{a_1}-\bm{a_2})\cdot\bm{k}},
\end{eqnarray}
\begin{eqnarray}
 H_{1.3}(\bm{k})=t_1+ t_1e^{-i\bm{a_2}\cdot\bm{k}}+t_2e^{-i\bm{a_1}\cdot\bm{k}}+t_2e^{i(\bm{a_1}-\bm{a_2})\cdot\bm{k}},
 \end{eqnarray}
\begin{eqnarray}
H_{2.1}(\bm{k})= t_1+ t_1e^{i\bm{a_1}\cdot\bm{k}}+t_2e^{i\bm{a_2}\cdot\bm{k}}+t_2e^{i(\bm{a_1}-\bm{a_2})\cdot\bm{k}},
\end{eqnarray} 
\begin{eqnarray}
H_{2.2}(\bm{k})= t_3e^{i\bm{a_2}\cdot\bm{k}}+ t_3e^{-i\bm{a_2}\cdot\bm{k}}+t_4e^{i\bm{a_1}\cdot\bm{k}}+t_4e^{-i\bm{a_1}\cdot\bm{k}}+t_4e^{i(\bm{a_1}-\bm{a_2})\cdot\bm{k}}+t_4e^{-i(\bm{a_1}-\bm{a_2})\cdot\bm{k}},
\end{eqnarray}  
\begin{eqnarray}
H_{2.3}(\bm{k})= t_1+ t_1e^{i(\bm{a_1}-\bm{a_2})\cdot\bm{k}}+t_2e^{-i\bm{a_2}\cdot\bm{k}}+t_2e^{i\bm{a_1}\cdot\bm{k}},
\end{eqnarray}  
 \begin{eqnarray}
H_{3.1}(\bm{k})= t_1+ t_1e^{i\bm{a_2}\cdot\bm{k}}+t_2e^{i\bm{a_1}\cdot\bm{k}}+t_2e^{-i(\bm{a_1}-\bm{a_2})\cdot\bm{k}},
\end{eqnarray} 
 \begin{eqnarray}
H_{3.2}(\bm{k})= t_1+ t_1e^{-i(\bm{a_1}-\bm{a_2})\cdot\bm{k}}+t_2e^{i\bm{a_2}\cdot\bm{k}}+t_2e^{-i\bm{a_1}\cdot\bm{k}},
\end{eqnarray} 
 \begin{eqnarray}
H_{3.3}(\bm{k})= t_3e^{i\bm{a_1}\cdot\bm{k}}+ t_3e^{-i\bm{a_1}\cdot\bm{k}}+t_4e^{i\bm{a_2}\cdot\bm{k}}+t_4e^{-i\bm{a_2}\cdot\bm{k}}+t_4e^{i(\bm{a_1}-\bm{a_2})\cdot\bm{k}}+t_4e^{-i(\bm{a_1}-\bm{a_2})\cdot\bm{k}}.
\end{eqnarray} 
\end{subequations}}

\textcolor{black}{It is to be noted that the $(i, j)$ components of $\mathcal{H}(\bm{k})$ denotes the electron hopping between the sublattices $i$ and $j$ in Fig.~\ref{fig:fig3}(a). In addition, $\bm{a_{1}}$ and $\bm{a_{2}}$ are the primitive translation vectors of the kagome lattice as shown in Fig.~\ref{fig:fig3}(a). In this study, we set $\bm{a_{1}}$ and $\bm{a_{2}}$ to $(a,0)$ and $(\frac{a}{2},\frac{\sqrt{3}a}{2})$, respectively, where $a$ is the length of $\bm{a_{1}}$ and $\bm{a_{2}}$. Since the coordinate of the $\rm{M}$ point is $\frac{2\pi}{a}(0,\frac{1}{\sqrt{3}})$ in the wave number space, and we get $\bm{k}=\frac{2k\pi}{a}(0,\frac{1}{\sqrt{3}})$ $(0\leqq k \leqq 1)$ when $\bm{k}$ is located on the $\Gamma-\rm{M}$ line. Therefore, we can obtain Eq.~\ref{eq:eqr1}.}

\section{Detailed derivation of the condition of $(t_1,t_2,t_3)$ for designing single flat band along $\Gamma-\rm{M}$ line}
\label{appendix:AppendixB}
In this section, we prove that $t_1+3t_2+2t_3=0$ holds when the highest band becomes flat along the $\Gamma-\rm{M}$ line.

\subsection{Necessary condition for single flat band along $\Gamma-\rm{M}$ line}
Firstly, we derive the necessary condition for the highest band to be flat along the $\Gamma-\rm{M}$ line.
The maximum eigenvalues of. $\mathcal{H}_{\rm{\Gamma}}$ and $\mathcal{H}_{\rm{M}}$ must coincide. 
We can easily obtain the maximum eigenvalue of $\mathcal{H}_{\rm{M}}$, $2t_1-2t_2-2t_3$. Next, let us consider the situation that $\mathcal{H}_{\rm{\Gamma}}$ also has the eigenvalue of $2t_1-2t_2-2t_3$:

\begin{equation}
\mathcal{H}_{\rm{\Gamma}}\bm{\psi}=(2t_1-2t_2-2t_3)\bm{\psi},
\label{eq:eqrS3}
\end{equation}

\begin{equation}
\bm{\psi}=\begin{pmatrix}
r_1\\
r_2\\
r_3\\
\end{pmatrix}.
\label{eq:eqrS4}
\end{equation}
Note that $\bm{\psi}\neq\bm{0}$. Eq.~(\ref{eq:eqrS3}) is a simultaneous ternary linear equation of $r_1$,$r_2$ and $r_3$ as follows:

\begin{subequations}
 \begin{eqnarray}
 (-2t_1+2t_2+4t_3)r_1+(2t_1+2t_2)r_2+(2t_1+2t_2)r_3=0,
 \end{eqnarray}
 \begin{eqnarray}
 (2t_1+2t_2)r_1+(-2t_1+2t_2+4t_3)r_2+(2t_1+2t_2)r_3=0,
 \end{eqnarray} 
 \begin{eqnarray}
(2t_1+2t_2)r_1+(2t_1+2t_2)r_2+(-2t_1+2t_2+4t_3)r_3=0.
\end{eqnarray}
\label{eq:eqrS5}
\end{subequations}
By adding both sides of these three equations together, we obtain

\begin{equation}
(2t_1+6t_2+4t_3)(r_1+r_2+r_3)=0.
\label{eq:eqrS6}
\end{equation}
Here, if we assume that $2t_1+6t_2+4t_3$ is not 0, $r_1+r_2+r_3=0$ is required. Under this condition, Eq.~\ref{eq:eqrS5}(a) can be represented as $(-4t_1+4t_3)r_1=0$. Due to $t_3\leqq0<t_1$, $r_1$ is 0. Similarly, $r_2=0$ is also obtained. Therefore, $\bm{r}=\bm{0}$ and this result is inappropriate. Hence, $t_1+3t_2+2t_3=0$ is required to have the eigenvalue of $2t_1-2t_2-2t_3$ for $\mathcal{H}_{\Gamma}$. The condition, $t_1+3t_2+2t_3=0$, is required for the highest band to be flat along the $\Gamma-\rm{M}$ line.

\subsection{Single flat band along $\Gamma-\rm{M}$ line}
Under the condition of $t_1+3t_2+2t_3=0$, it is found that $\mathcal{H}(k)$ has the maximum eigenvalue of $2t_1-2t_2-2t_3$ regardless of the $k$ value ($0 \leqq k \leqq1$):

\begin{equation}
\mathcal{H}(k)\bm{\psi}=(2t_1-2t_2-2t_3)\bm{\psi},
\label{eq:eqrS11}
\end{equation}
where
\begin{equation}
\bm{\psi}=
\begin{pmatrix}
1 \\
1 \\
\frac{e^{ik\pi}-\cos(k\pi)}{1-e^{-ik\pi}} \\
\end{pmatrix}. \\
\label{eq:eqrS12}
\end{equation}
This indicates that the highest band becomes flat along the $\Gamma-\rm{M}$ line when $t_1+3t_2+2t_3=0$ holds, as shown in Fig.~\ref{fig:fig1}(a).

\section{Detailed derivation of the condition of $(t_1,t_2,t_3)$ for designing degenerate flat band along $\Gamma-\rm{M}$ line}
\label{appendix:AppendixB2}

We show that $t_1=-t_2=-t_3$ holds when the second and the third highest bands degenerate, and they are flat along the $\Gamma-\rm{M}$ line.

\subsection{Necessary condition for degenerate band between $\Gamma-\rm{M}$ line}
Firstly, we derive the necessary condition for the second and the third highest bands to be degenerate along the $\Gamma-\rm{M}$ line. The first minimum and the second minimum eigenvalues of $\mathcal{H}_{\rm{M}}$, $-2t_1+2t_2-2t_3$ and $2t_3$, must coincide. Hence, $2t_3=-t_1+t_2$ is required. In the discussion that follows, this equation is assumed to hold. Next, $\mathcal{H}_{\rm{middle}}$ must have degenerate eigenvalues, too. We can obtain the eigenvalues of $\mathcal{H}_{\rm{middle}}$ by solving the cubiq equation for $\lambda$:

\begin{equation}
\rm{det}\begin{pmatrix}
-\lambda & 2t_1 & (t_1+t_2)(1-i)\\
2t_1& -\lambda & (t_1+t_2)(1-i)\\
(t_1+t_2)(1+i) & (t_1+t_2)(1+i) & 2t_3-\lambda \\
\end{pmatrix}
=0
\label{eq:eqrS7}
\end{equation}
This equation can be transformed into

\begin{equation} 
(\lambda+2t_1)(\lambda^2-(2t_1+2t_3)\lambda+(4t_1t_3-4(t_1+t_2)^2))=0.
\label{eq:eqrS8}
\end{equation}
For $\mathcal{H}_{\rm{middle}}$ to have degenerate solution, the quadratic equation $\lambda^2-(2t_1+2t_3)\lambda+(4t_1t_3-4(t_1+t_2)^2)=0$ must have (A) solution of $\lambda=-2t_1$ or (B) doubly degenerate solutions. \\\\\textbf{Case of (A)}: This quadratic equation can be rewritten as ${t_1}^2-2(t_2-t_3)t_1-{t_2}^2=0$. By substituting $2t_3=-t_1+t_2$ for it, we can obtain $2t_2t_3=0$, indicating that $t_2=0$ or $t_3=0$ must hold. If $t_3$ is 0, $t_1=-t_2$ can be obtained from $2t_3=-t_1+t_2$, and then we can transform the equation ${t_1}^2-2(t_2-t_3)t_1-{t_2}^2=0$ into $2{t_1}^2=0$. Hence, $t_1=t_2=t_3=0$ holds and this equation is inappropriate. This discussion indicates that $t_2=0$ and $t_1=-2t_3$ is necessary in the case of (A).
\\\\\textbf{Case of (B)}: The equation, $(t_1+t_3)^2-(4t_1t_3-4(t_1+t_2)^2)=0$, is required by considering discriminant. This equation can be transformed into $(t_1-t_3)^2=-4(t_1+t_2)^2$ and therefore, $t_1=-t_2=-t_3 (\neq 0)$ is required.
\\\\From these discussions, (i) $t_2=0$ and $t_1+2t_3=0$, or (ii) $t_1=-t_2=-t_3 (\neq 0)$ is required when the second and the third highest bands are degenerate along the $\Gamma-\rm{M}$ line.

\subsection{Cosine-type degenerate band and flat degenerate band between $\Gamma-\rm{M}$ line}

Next, we solve the eigenvalue problem of $\mathcal{H}(k)$ under the condition of (i) $t_2=0$ and $t_1+2t_3=0$, and (ii) $t_1=-t_2=-t_3 (\neq 0)$, respectively. \\\\ \textbf{(i) \bm{$t_2=0$} and \bm{$t_1+2t_3=0$}}: the eigenvalues of $\mathcal{H}(k)$ are $-t_1(2+\cos(k\pi))$, $-t_1(2+\cos(k\pi))$ and $3t_1$, indicating that the cosine-type degenerate band along the $\Gamma-\rm{M}$ line appears as shown in Fig.~\ref{fig:fig1}(c1).
\\\\\textbf{(ii) \bm{$t_1=-t_2=-t_3(\neq0)$}}: The eigenvalues of $\mathcal{H}(k)$ are $-2t_1$, $-2t_1$ and $t_1(-4\cos(k\pi)+2)$, indicating that the degenerate flat band along the $\Gamma-\rm{M}$ line appears as shown in Fig.~\ref{fig:fig1}(b).

\section{Detailed derivation of the condition of $(t_1,t_2,t_3,t_4)$ for designing degenerate band along $\Gamma-\rm{M}$ line}
\label{appendix:AppendixC}
\textcolor{black}{In this section, we show that (i) $t_1+2t_3-2t_4=0$ and $t_2=0$, or (ii) $t_1=-t_2$ and $t_4-t_3=t_1$ holds when the second highest and the third highest bands are degenerate along the $\Gamma-\rm{M}$ line.}

\subsection{Necessary condition for degenerate band along $\Gamma-\rm{M}$ line}\label{appendix:AppendixC1}
\textcolor{black}{
We derive the necessary condition for the second and the third highest bands to be degenerate along the $\Gamma-\rm{M}$ line. The first minimum and the second minimum eigenvalues of $\mathcal{H}_{\rm{M}}$, $-2t_1+2t_2-2t_3$ and $2t_3-4t_4$, must coincide. Hence, $t_1-t_2+2t_3-2t_4=0$ is required. In the discussion that follows, this equation is assumed to hold. Next, the first minimum and the second minimum eigenvalues of $\mathcal{H}_{\rm{middle}}$ must coincide, too. We can obtain the eigenvalues of $\mathcal{H}_{\rm{middle}}$ by solving the cubiq equation for $\lambda$:}

\textcolor{black}{
\begin{equation}
\rm{det}\begin{pmatrix}
2t_4-\lambda & 2t_1 & (t_1+t_2)(1-i)\\
2t_1& 2t_4-\lambda & (t_1+t_2)(1-i)\\
(t_1+t_2)(1+i) & (t_1+t_2)(1+i) & 2t_3-\lambda \\
\end{pmatrix}
=0
\label{eq:eqrS12}
\end{equation}
This equation can be transformed into}

\textcolor{black}{
\begin{equation} 
\{\lambda-(2t_4-2t_1)\}[\lambda^2-2(t_1+t_3+t_4)\lambda+\{4t_3(t_1+t_4)-4(t_1+t_2)^2\}]=0.
\label{eq:eqrS13}
\end{equation}
Therefore, we can obtain three solutions,}

\textcolor{black}{
\begin{subequations}
\label{eq:eqApC1}
\begin{eqnarray}
\lambda_{1}=2t_4-2t_1,
\end{eqnarray}
\begin{eqnarray}
\lambda_{2}=(t_1+t_3+t_4)-\sqrt{(t_1+t_3+t_4)^2-4t_3(t_1+t_4)+4(t_1+t_2)^2},
\end{eqnarray}
\begin{eqnarray}
\lambda_{3}=(t_1+t_3+t_4)+\sqrt{(t_1+t_3+t_4)^2-4t_3(t_1+t_4)+4(t_1+t_2)^2}.
\end{eqnarray}
\end{subequations}
Due to $t_3\leqq0$,  the value of $\sqrt{(t_1+t_3+t_4)^2-4t_3(t_1+t_4)+4(t_1+t_2)^2}$ is larger than 0. Therefore, $\lambda_{2}<\lambda_{3}$ holds and then $\lambda_{1}=\lambda_{2}$ is necessary. The equation, $\lambda_{1}=\lambda_{2}$, can be rewritten as:}

\textcolor{black}{
\begin{equation} 
t_4-3t_1-t_3=-\sqrt{(t_1+t_3+t_4)^2-4t_3(t_1+t_4)+4(t_1+t_2)^2}\\
\to (t_4-3t_1-t_3)^2 = (t_1+t_3+t_4)^2-4t_3(t_1+t_4)+4(t_1+t_2)^2
\label{eq:eqrS14}
\end{equation}
By substituting $t_1-t_2+2t_3-2t_4=0$ for this, we can obtain}

\textcolor{black}{
\begin{equation} 
(t_1+2t_3-2t_4)(t_1+t_3-t_4)=0
\label{eq:eqrS15}.
\end{equation}
This equation indicates that (i) $t_1+2t_3-2t_4=0$ or (ii) $t_1=-t_2$ and $t_4-t_3=t_1$ is required when the second and the third highest bands are degenerate along the $\Gamma-\rm{M}$ line.}

\subsection{Eigen energy under the condition of $t_1+2t_3-2t_4=0$ and $t_2=0$}
\label{appendix:AppendixC2}
\textcolor{black}{Let us consider the situation which satisfies $t_1+2t_3-2t_4=0$ and $t_2=0$. Firstly, when $t_2=0$, $\mathcal{H}(k)$ can be expressed as}

\textcolor{black}{
\begin{equation}
\mathcal{H}(k)=\begin{pmatrix}
2t_3\cos(k\pi)+2t_4[1+\cos(k\pi)] & 2t_1& t_1(1+e^{-ik\pi})\\
2t_1 &  2t_3\cos(k\pi)+2t_4[1+\cos(k\pi)] & t_1(1+e^{-ik\pi}) \\
t_1(1+e^{ik\pi}) & t_1(1+e^{ik\pi}) & 2t_3+4t_4\cos(k\pi) \\
\end{pmatrix}.
\label{eq:eqrS9}
\end{equation}
Next, by using $(t_3, t_4) =(\alpha -\frac{1}{2}, \alpha)t_1$ $(0\leq\alpha)$, we can obtain} 

\textcolor{black}{
\begin{equation}
\mathcal{H}(k)=\begin{pmatrix}
2\alpha+(4\alpha-1)\cos(k\pi)& 2& 1+e^{-ik\pi}\\
2 & 2\alpha+(4\alpha-1)\cos(k\pi) & 1+e^{-ik\pi} \\
1+e^{ik\pi} & 1+e^{ik\pi} & (2\alpha-1)+4\alpha\cos(k\pi) \\
\end{pmatrix}t_1.
\label{eq:eqrS10}
\end{equation}
Therefore, the eigenequation of $\mathcal{H}(k)$ is $\{\lambda-[(2\alpha-2)+(4\alpha-1)\cos(k\pi)]t_1\}^2\{\lambda-[(2\alpha+3)+4\alpha\cos(k\pi)]t_1\}=0$ where $\lambda$ is the eigenvalue, and we find that $\mathcal{H}(k)$ has doubly degenerate eigenvalue, $\lambda_{\rm{deg}}(k)=[(2\alpha-2)+(4\alpha-1)\cos(k\pi)]t_1$, and non-degenerate eigenvalue, $\lambda_{\rm{non-deg}}(k)=[(2\alpha+3)+4\alpha\cos(k\pi)]t_1$. }

\subsection{Eigen energy under the condition of $t_1=-t_2$ and $t_4-t_3=t_1$}
\textcolor{black}{Let us consider the situation which satisfies $t_1=-t_2$ and $t_4-t_3=t_1$. Firstly, due to $t_1=-t_2$, $\mathcal{H}(k)$ can be expressed as}

\textcolor{black}{
\begin{equation}
\mathcal{H}(k)=\begin{pmatrix}
2t_3\cos(k\pi)+2t_4[1+\cos(k\pi)] & 2t_1(1-\cos(k\pi)) & 0\\
2t_1(1-\cos(k\pi)) &  2t_3\cos(k\pi)+2t_4[1+\cos(k\pi)] & 0 \\
0 & 0 & 2t_3+4t_4\cos(k\pi) \\
\end{pmatrix}.
\label{eq:eqrS16}
\end{equation}
Therefore, we get doubly degenerate eigenvalue, $\lambda_{\rm{deg}}(k)=\{(2\alpha-2)+4\alpha\cos(k\pi)\}t_1$ and non-degenerate eigenvalue, $\lambda_{\rm{non-deg}}(k)=\{(2\alpha+2)+(4\alpha-4)\cos(k\pi)\}t_1$, where $(t_3, t_4)=(\alpha-1,\alpha)t_1$ $(0\leq\alpha\leq1)$. In Fig.~\ref{fig:figAppendixD}, we show the electronic structures. }

\begin{figure*}
\includegraphics[scale=0.85]{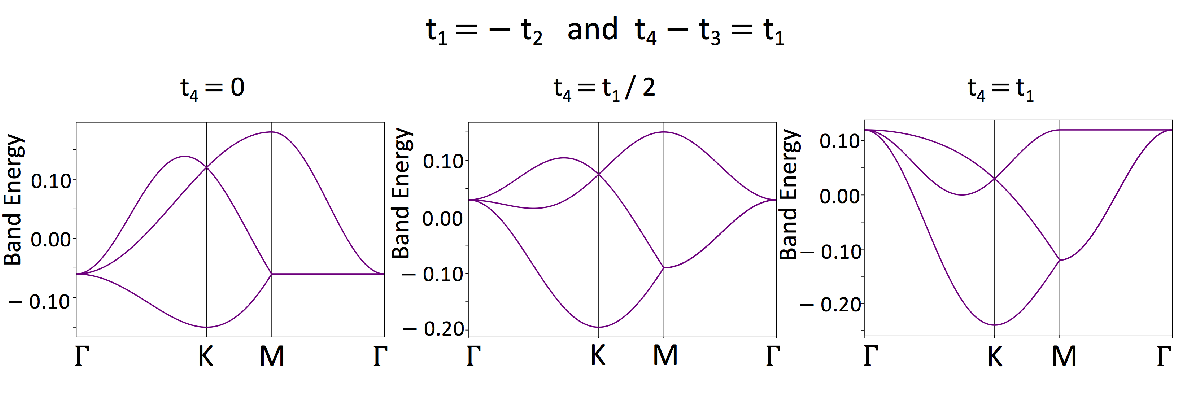}
\caption{\label{fig:figAppendixD} Electronic structures of kagome lattice when $(t_1, t_2, t_3, t_4)$ satisfies $t_1=-t_2$ and $t_4-t_3=t_1$. The electronic structure of $t_4=0$ is same as Fig.~\ref{fig:fig1}(b).}
\end{figure*}

\section{Nearly degenerate band in the vicinity of $\Gamma$ point under the condition of $t_1+2t_3-2t_4=0$ and $t_2=0$}\label{appendix:AppendixD}
\textcolor{black}{In this section, we prove the band degeneracy approximately occurs in the vicinity of $\Gamma$ point ($\bm{k}\simeq\bm{0}$) under the condition of $t_1+2t_3-2t_4=0$ and $t_2=0$.} 

\textcolor{black}{Firstly, let us define $\mathcal{H}^{\rm{vic\Gamma}}(\bm{k})$ as the hamiltonian obtained by considering up to the second order terms of $\bm{k}$ in $\mathcal{H}(\bm{k})$ (Eq.~\ref{eq:eqr_hamiltonian}). In particular, when $t_2=0$, we can get}

\textcolor{black}{
\begin{equation}
\mathcal{H}^{\rm{vic \Gamma}}(\bm{k})=\begin{pmatrix}
H_{1.1}^{\rm{vic \Gamma}}(\bm{k}) & H_{1.2}^{\rm{vic \Gamma}}(\bm{k})& H_{1.3}^{\rm{vic \Gamma}}(\bm{k})\\
H_{2.1}^{\rm{vic \Gamma}}(\bm{k}) & H_{2.2}^{\rm{vic \Gamma}}(\bm{k}) & H_{2.3}^{\rm{vic \Gamma}}(\bm{k}) \\
H_{3.1}^{\rm{vic \Gamma}}(\bm{k}) & H_{3.2}^{\rm{vic \Gamma}}(\bm{k}) & H_{3.3}^{\rm{vic \Gamma}}(\bm{k}) \\
\end{pmatrix}
\label{eq:eqr10}
\end{equation}
where
\begin{subequations}
\label{eqr11a}
\begin{eqnarray}
H_{1.1}^{\rm{vic \Gamma}}(\bm{k})= t_3[2-((\bm{a_1}-\bm{a_2})\cdot\bm{k})^2]+t_4[4-(\bm{a_1}\cdot\bm{k})^2-(\bm{a_2}\cdot\bm{k})^2],
\end{eqnarray}
\begin{eqnarray}
H_{1.2}^{\rm{vic \Gamma}}(\bm{k})= t_1\Big[2-i\bm{a_1}\cdot\bm{k}-\frac{1}{2}(\bm{a_1}\cdot\bm{k})^2\Big],
\end{eqnarray}
\begin{eqnarray}
 H_{1.3}^{\rm{vic \Gamma}}(\bm{k})= t_1\Big[2-i\bm{a_2}\cdot\bm{k}-\frac{1}{2}(\bm{a_2}\cdot\bm{k})^2\Big],
 \end{eqnarray}
\begin{eqnarray}
H_{2.1}^{\rm{vic \Gamma}}(\bm{k})= t_1\Big[2+i\bm{a_1}\cdot\bm{k}-\frac{1}{2}(\bm{a_1}\cdot\bm{k})^2\Big],
\end{eqnarray} 
\begin{eqnarray}
H_{2.2}^{\rm{vic \Gamma}}(\bm{k})= t_3[2-(\bm{a}_{2}\cdot\bm{k})^2]+t_4[4-(\bm{a}_{1}\cdot\bm{k})^2-((\bm{a_1}-\bm{a_2})\cdot\bm{k})^2],
\end{eqnarray}
  \begin{eqnarray}
H_{2.3}^{\rm{vic \Gamma}}(\bm{k})= t_1\Big[2+i(\bm{a_1}-\bm{a_2})\cdot\bm{k}-\frac{1}{2}((\bm{a_1}-\bm{a_2})\cdot\bm{k})^2\Big],
\end{eqnarray}  
 \begin{eqnarray}
H_{3.1}^{\rm{vic \Gamma}}(\bm{k})= t_1\Big[2+i\bm{a_2}\cdot\bm{k}-\frac{1}{2}(\bm{a_2}\cdot\bm{k})^2\Big],
\end{eqnarray} 
 \begin{eqnarray}
H_{3.2}^{\rm{vic \Gamma}}(\bm{k})= t_1\Big[2-i(\bm{a_1}-\bm{a_2})\cdot\bm{k}-\frac{1}{2}((\bm{a_1}-\bm{a_2})\cdot\bm{k})^2\Big],
\end{eqnarray} 
\begin{eqnarray}
H_{3.3}^{\rm{vic \Gamma}}(\bm{k})= t_3[2-(\bm{a}_{1}\cdot\bm{k})^2]+t_4[4-(\bm{a}_{2}\cdot\bm{k})^2-((\bm{a_1}-\bm{a_2})\cdot\bm{k})^2].
\end{eqnarray}
 \end{subequations}
Next, by using $(t_3,t_4)=(\alpha-\frac{1}{2},\alpha)t_1$ ($0\leq \alpha$), $H_{1,1}^{\rm{vic\Gamma}}(\bm{k})$, $H_{2,2}^{\rm{vic\Gamma}}(\bm{k})$ and $H_{3,3}^{\rm{vic\Gamma}}(\bm{k})$ can be rewritten as}

\textcolor{black}{
\begin{subequations}
 \begin{eqnarray}
 H_{1.1}^{\rm{vic \Gamma}}(\bm{k})= t_1\Big[(6\alpha-1)-\alpha[(\bm{a_1}\cdot\bm{k})^2+(\bm{a_2}\cdot\bm{k})^2]-\Big(\alpha-\frac{1}{2}\Big)((\bm{a_1}-\bm{a_2})\cdot\bm{k})^2\Big],
 \end{eqnarray}
 \begin{eqnarray}
 H_{2.2}^{\rm{vic \Gamma}}(\bm{k})= t_1\Big[(6\alpha-1)-\alpha[(\bm{a_1}\cdot\bm{k})^2+((\bm{a_1}-\bm{a_2})\cdot\bm{k})^2]-\Big(\alpha-\frac{1}{2}\Big)(\bm{a_2}\cdot\bm{k})^2\Big],
 \end{eqnarray} 
 \begin{eqnarray}
 H_{3.3}^{\rm{vic \Gamma}}(\bm{k})= t_1\Big[(6\alpha-1)-\alpha[(\bm{a_2}\cdot\bm{k})^2+((\bm{a_1}-\bm{a_2})\cdot\bm{k})^2]-\Big(\alpha-\frac{1}{2}\Big)(\bm{a_1}\cdot\bm{k})^2\Big].
\end{eqnarray}
\label{eq:eqrS12}
\end{subequations}
Therefore, we get the eigen equation of $\mathcal{H}^{\rm{vic\Gamma}}(\bm{k})/{t_1}-(6\alpha-1)I_{3}$ ($I_{3}$ is 3$\times$3 unit matrix) as follows:}

\textcolor{black}{
\begin{equation}
\lambda^3+a\lambda^2+b\lambda+c + o(\bm{k}^2) (\bm{k}^2 \rightarrow 0) = 0
\label{eq:eqr13}
\end{equation}
where
\begin{subequations}
\label{eqr14a}
\begin{eqnarray}
a = (6\alpha-1)[(\bm{a_1}\cdot\bm{k})^2+(\bm{a_2}\cdot\bm{k})^2-(\bm{a_1}\cdot\bm{k})(\bm{a_2}\cdot\bm{k})],
\end{eqnarray}
\begin{eqnarray}
b = 2[(\bm{a_1}\cdot\bm{k})^2+(\bm{a_2}\cdot\bm{k})^2-(\bm{a_1}\cdot\bm{k})(\bm{a_2}\cdot\bm{k})]-12,
\end{eqnarray}
\begin{eqnarray}
c =(24\alpha-8)[-(\bm{a_1}\cdot\bm{k})^2-(\bm{a_2}\cdot\bm{k})^2+(\bm{a_1}\cdot\bm{k})(\bm{a_2}\cdot\bm{k})]-16.
 \end{eqnarray}
 \end{subequations}
When $o(\bm{k}^2) (\bm{k}^2 \rightarrow 0)$ is ignored, this equation has the doubly degenerate eigenvalue $\lambda_{\rm{deg}}^{\rm{vic\Gamma}}(\bm{k})=-2-\frac{4\alpha-1}{2}((\bm{a_1}+\bm{a_2})\cdot\bm{k})^2+\Big(6\alpha-\frac{3}{2}\Big)(\bm{a_1}\cdot\bm{k})(\bm{a_2}\cdot\bm{k})$ and the non degenerate eigenvalue $\lambda_{\rm{non-deg}}^{\rm{vic\Gamma}}(\bm{k})=4-2\alpha((\bm{a_1}+\bm{a_2})\cdot\bm{k})^2+6\alpha(\bm{a_1}\cdot\bm{k})(\bm{a_2}\cdot\bm{k})$. Hence, in the vicinity of $\Gamma$ point ($\bm{k}\simeq\bm{0}$), the second highest and the third highest band almost degenerate regardless of the $\bm{k}$ direction. In particular, when $\alpha=\frac{1}{4}$ (Fig.~\ref{fig:fig1}(c3)), the value of $\lambda_{\rm{deg}}^{\rm{vic\Gamma}}(\bm{k})$ is $-2$, indicating that the degenerate band is flat in the vicinity of $\Gamma$ point.}

\section{Effect of small perpendicular hopping on electronic structure of kagome lattice}\label{appendix:AppendixE}
In this section, we discuss how the flat band is modified by small perpendicular hopping. To discuss it, we consider $AA$ and $AB$ stacking cases of the kagome lattice as shown below.

\subsection{Setup}
In this work, a two dimensional kagome layer, where hopping satisfies $(t_1,t_2,t_3,t_4)=(1,0,-\frac{1}{4},\frac{1}{4})t_1$ (Fig.~\ref{fig:fig1}(c3)), is considered. We study the systems where these layers are stacked infinitely. By introducing small perpendicular hopping between different layers, quasi two-dimensional kagome lattice can be constructed. Small perpendicular hopping is only allowed between sublattices, which have same $x$ and $y$ coordinates.  This hopping value does not depend on the combination of sublattices. 

\subsection{Results}
\subsubsection{$\textit{AA}$ stacking kagome layers}
Firstly, we discuss the electronic structure of $\textit{AA}$ stacking kagome layers. Therefore, the perpendicular hopping is only allowed between the same sublattices. The Bloch hamiltonian of \textit{AA} stacking layers (3$\times$3 matrix)  is the sum of $\mathcal{H}(\bm{k})$ (Eq.~\ref{eq:eqr_hamiltonian}) and the diagonal matrix derived from the perpendicular hopping. Since all the diagonal components of the latter matrix are same, the electronic structure is same as Fig.~\ref{fig:fig1}(c3) in the main text, regardless of the $k_z$ value. 

\subsubsection{$\textit{AB}$ stacking kagome layers}
Next, we discuss the electronic structure of $\textit{AB}$ stacking kagome layers. The crystal structure and the electronic structure are shown in Fig.~\ref{fig:figSprb}. In the unit cell, there are two kagome layers and six sublattices (1,2,3,4,5,6) as shown in Fig.~\ref{fig:figSprb}(a). It is found that the degenerate flat band along the $\Gamma-\rm{M}$ line is maintained at $k_{z}=\frac{\pi}{c}$ while it is not maintained at $k_{z}=0$ after stacking.

\begin{figure}[h]
\includegraphics[scale=0.65]{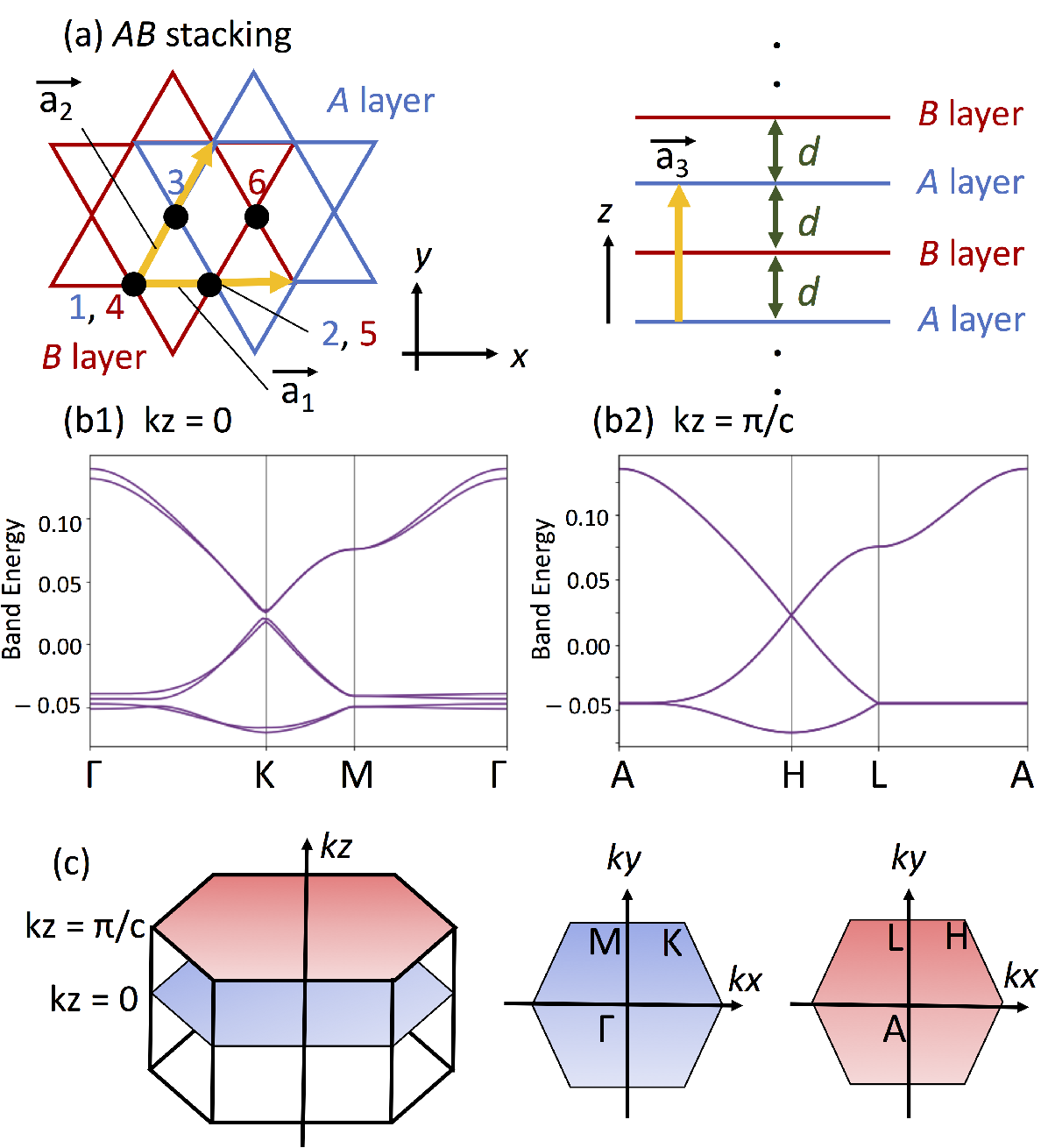}
\caption{\label{fig:figSprb} (a) Schematic views and electronic structures at (b1) $k_{z}=0$ and (b1) $k_{z}=\frac{\pi}{c}$ of \textit{AB} stacking kagome layers. $A$ kagome layer and $B$ kagome layer are stacked alternately. In (a), $\vec{a_{1}}=(a,0,0)$, $\vec{a_{2}}=(\frac{a}{2},\frac{\sqrt{3}a}{2},0)$ and $\vec{a_{3}}=(0,0,c)$ represent the primitive translation vectors. Three sublattices of \textit{A} layer (1,2,3) and three sublattices of \textit{B} layer (4,5,6) are also shown.  Between different layers, small perpendicular hopping ($d=\frac{t_1}{10}$) is introduced. (c) First Brillouin zone of kagome layers.}
\end{figure}

This origin can be revealed analytically by considering the Bloch hamiltonian. By using $\mathcal{H}(\bm{k})$ (Eq.~\ref{eq:eqr_hamiltonian}) and $\mathcal{H}_{\rm{per}}(\bm{k})$ derived from the perpendicular hopping, the Bloch hamiltonian of \textit{AB} stacking kagome layers , $\mathcal{H}_{AB}(\bm{k})$ (6$\times$6 matrix), can be expressed as

\begin{equation}
\mathcal{H}_{AB}(\bm{k})=\begin{pmatrix}
\mathcal{H}(\bm{k}) & \mathcal{H}_{\rm{per}}(\bm{k})\\
\mathcal{H}^{\dagger}_{\rm{per}}(\bm{k}) & \mathcal{H}(\bm{k})\\
\end{pmatrix}.
\label{eq:eqrprb1}
\end{equation}
It is to be noted that the $(i,j)$ component of $\mathcal{H}_{AB}(\bm{k})$ ($1\leq i,j \leq6$) denotes the electron hopping between the sublattices $i$ and $j$ shown in Fig.~\ref{fig:figSprb}(a). 

Actually, we can easily prove that $\mathcal{H}_{\rm{per}}(\bm{k})=\Huge{0}$. Let us focus on the hopping between sublattices 1 and 4, which is one of the components of $\mathcal{H}_{\rm{per}}(\bm{k})$. From Fig.~\ref{fig:figSprb}(a), this hopping is $d+de^{i\bm{a}_{3}\cdot\bm{k}}$, where $\bm{a}_{3}=(0,0,c)$ and $\bm{k}=(k_x,k_y,k_z)$. Therefore, in the case of $k_z=\frac{\pi}{c}$, the hopping between the sublattices 1 and 4 is 0. Similarly, the hopping between the sublattices 2 and 5 is 0. Therefore, $\mathcal{H}_{\rm{per}}(\bm{k})=\Huge{0}$, and the degenerate flat band can be maintained.

\section{Calculation methods for $(8,L)$ graphene}
\label{appendix:Appendix_method}

\subsection{Electronic structure}
\label{appendix:Appendix_method_DFT}
\textcolor{black}{We performed first-principles total-energy calculations using the generalized gradient approximation (GGA) within the framework of the density functional theory (DFT) \cite{hohenberg1964inhomogeneous, kohn1965self} to obtain the electronic structure of $(8,L)$ graphene. We used the energy functional by Perdew, Burke, and Ernzerhof together with the Vanderbilt ultrasoft pseudipotentials to describe the interactions between the ions and valence electrons \cite{perdew1996generalized, perdew78errata, vanderbilt1990soft}. The cutoff energies of the plene-wave basis and the charge density are taken to be 60 Ry and 480 Ry, respectively. The interlayer distance between graphene was kept to be greater than 10 $\rm{\AA}$ to simulate isolated graphene sheets. }

\subsection{Maximally localized Wannier function reproducing the first, the second and the third LU bands}
\label{appendix:Appendix_MLWF}
\textcolor{black}{We constructed the maximally localized Wannier functions (MLWFs) and the tight binding models which reproduce the first, the second and the third LU bands of $(8,L)$ graphene obtained by the GGA calculation. As the initial MLWFs, we set three $p_z$ orbitals at the middle of subribbon part as shown in Fig.~\ref{fig:figS2} (a).}

\subsection{Optical conductivity}
\label{appendix:Appendix_optical}
\textcolor{black}{We constructed the MLWFs and the tight binding model which reproduce the first, the second and the third LU and HO bands of $(8,L)$ graphene obtained by the GGA calculation. As the initial MLWFs, we set six $p_{z}$ orbitals at the middle of zigzag edge.}

\textcolor{black}{Next, we calculate the optical conductivity based on these six MLWFs. The optical conductivity can be obtained from the Kubo-Greenwood formula:}

\textcolor{black}{
\begin{equation}
\sigma_{\alpha\beta}(\hbar\omega)=\frac{ie^2\hbar}{N_{k}\Omega_{C}}\sum_{\bm{k}}\sum_{n,m}\frac{f_{m\bm{k}}-f_{n\bm{k}}}{\epsilon_{m\bm{k}}-\epsilon_{n\bm{k}}}\frac{\langle\psi_{n\bm{k}}|v_{\alpha}|\psi_{m\bm{k}}\rangle\langle\psi_{m\bm{k}}|v_{\beta}|\psi_{n\bm{k}}\rangle}{\epsilon_{m\bm{k}}-\epsilon_{n\bm{k}}-(\hbar\omega+i\eta)}
\end{equation}
It is to be noted that $\alpha$ and $\beta$ are Cartesian directions, $\Omega_{C}$ is the cell volume, $N_{k}$ is the number of $k$-points used for sampling in the Brillouin zone, $\epsilon_{m\bm{k}}$ is the band energy, and $f_{n\bm{k}}=f(\epsilon_{n\bm{k}})$ is the Fermi-Dirac distribution functions. By using $\langle\psi_{n\bm{k}}|v|\psi_{m\bm{k}}\rangle=-\frac{i}{\hbar}(\epsilon_{m\bm{k}}-\epsilon_{n\bm{k}})\bm{A}_{nm}(\bm{k}) (m\neq n)$, This optical conductivity can be rewritten as follows:}

\textcolor{black}{
\begin{equation}
\sigma_{\alpha\beta}(\hbar\omega)=\frac{ie^2}{N_{k}\hbar\Omega_{C}}\sum_{\bm{k}}\sum_{n,m}(f_{m\bm{k}}-f_{n\bm{k}})\frac{\epsilon_{m\bm{k}}-\epsilon_{n\bm{k}}}{\epsilon_{m\bm{k}}-\epsilon_{n\bm{k}}-(\hbar\omega+i\eta)}A_{nm,\alpha}(\bm{k})A_{mn,\beta}(\bm{k}).
\end{equation}
Here, $\bm{A}_{nm}(\bm{k})$ represents the connection matrix \cite{blount1962formalisms}. Next, we can decompose it into Hermitian part ($\sigma_{\alpha\beta}^{H}$) and anti-Hermitian ($\sigma_{\alpha\beta}^{AH}$) parts. In this paper, we show the real part of $\sigma_{\alpha\beta}^{H}$ ($\alpha\beta=xx$ $(yy)$). The $\sigma_{\alpha\beta}^{H}$ can be expressed as:}

\textcolor{black}{
\begin{equation}
\sigma_{\alpha\beta}^{H}(\hbar\omega)=-\frac{\pi e^2}{N_{k}\hbar\Omega_{C}}\sum_{\bm{k}}\sum_{n,m}(f_{m\bm{k}}-f_{n\bm{k}})(\epsilon_{m\bm{k}}-\epsilon_{n\bm{k}})A_{nm,\alpha}(\bm{k})A_{mn,\beta}(\bm{k})\bar{\delta}(\epsilon_{m\bm{k}}-\epsilon_{n\bm{k}}-\hbar\omega)
\end{equation}
where $\bar{\delta}(\epsilon)$ is a broadened delta-function. In the calculations, $144\times144\times144$ k mesh was applied, the convergence of which had been tested. }


\section{Calculation results of maximally localized Wannier function of $(8,16)$ graphene}\label{appendix:AppendixF}

\begin{figure}
\includegraphics[scale=0.80]{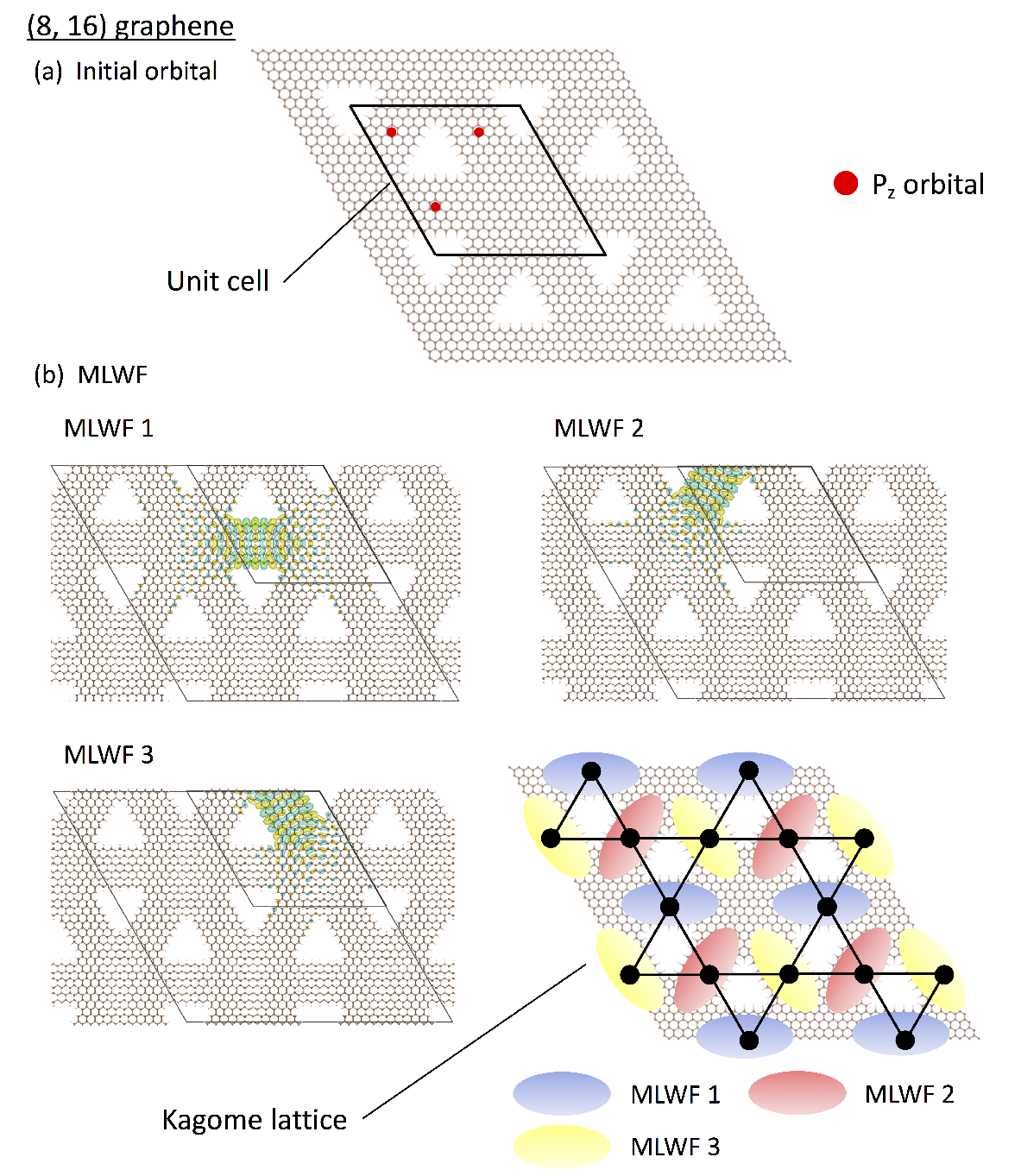}
\caption{\label{fig:figS2} (a) Initial orbitals for calculating MLWFs and (b) Three MLWFs of $(8,16)$ graphene. It is to be noted that sign differences of the MLWFs are shown by two colors.}
\end{figure}

The MLWFs (MLWF1, MLWF2 and MLWF3) in the case of $(8,16)$ graphene are also shown in Fig.~\ref{fig:figS2} (b).

\end{widetext}


\providecommand{\noopsort}[1]{}\providecommand{\singleletter}[1]{#1}%
%


\end{document}